\documentclass[superscriptaddress,prc,aps,twocolumn]{revtex4}
\usepackage{amsmath,amssymb}
\usepackage{graphicx}
\newcommand{\Mainz}[1]
{\affiliation{Institut f\"ur Kernphysik, University of Mainz, D-55099 Mainz,Germany}}
\newcommand{\Bonn}[1]
{\affiliation{Helmholtz-Institut f\"ur Strahlen- und Kernphysik, University of Bonn,
 D-53115 Bonn, Germany}}
\newcommand{\Regina}[1]
{\affiliation{University of Regina, Regina, Saskatchewan S4S 0A2, Canada}}
\newcommand{\Glasgow}[1]
{\affiliation{SUPA School of Physics and Astronomy, University of Glasgow,
 Glasgow G12 8QQ, United Kingdom}}
\newcommand{\Kent}[1]
{\affiliation{Kent State University, Kent, Ohio 44242-0001, USA}}
\newcommand{\Giessen}[1]
{\affiliation{II Physikalisches Institut, University of Giessen, D-3539 Giessen, Germany}}
\newcommand{\Dubna}[1]
{\affiliation{Joint Institute for Nuclear Research, 141980 Dubna, Russia}}
\newcommand{\Pavia}[1]
{\affiliation{INFN Sezione di Pavia, I-27100 Pavia, Italy}}
\newcommand{\GWU}[1]
{\affiliation{The George Washington University, Washington, DC 20052-0001, USA}}
\newcommand{\LPI}[1]
{\affiliation{Lebedev Physical Institute, 119991 Moscow, Russia}}
\newcommand{\Dalhousie}[1]
{\affiliation{Dalhousie University, Halifax, Nova Scotia B3H 4R2, Canada}}
\newcommand{\Halifax}[1]
{\affiliation{Saint Mary’s University, Halifax, Nova Scotia B3H 3C3, Canada}}
\newcommand{\UniPavia}[1]
{\affiliation{Dipartimento di Fisica, Universit\`a di Pavia, I-27100 Pavia, Italy}}
\newcommand{\Basel}[1]
{\affiliation{Institut f\"ur Physik, University of Basel, CH-4056 Basel, Switzerland}}
\newcommand{\Tomsk}[1]
{\affiliation{Laboratory of Mathematical Physics, Tomsk Polytechnic University, 634034 
Tomsk, Russia}}
\newcommand{\Edinburgh}[1]
{\affiliation{School of Physics, University of Edinburgh, Edinburgh EH9 3JZ,
 United Kingdom}}
\newcommand{\INR}[1]
{\affiliation{Institute for Nuclear Research, 125047 Moscow, Russia}}
\newcommand{\Sackville}[1]
{\affiliation{Mount Allison University, Sackville, New Brunswick E4L 1E6, Canada}}
\newcommand{\Zagreb}[1]
{\affiliation{Rudjer Boskovic Institute, HR-10000 Zagreb, Croatia}}
\newcommand{\Amherst}[1]
{\affiliation{University of Massachusetts, Amherst, Massachusetts 01003, USA}}
\newcommand{\UCLA}[1]
{\affiliation{University of California Los Angeles, Los Angeles, California 90095-1547, USA}}
\newcommand{\Jerusalem}[1]
{\affiliation{Racah Institute of Physics, Hebrew University of Jerusalem, Jerusalem 91904, Israel}}
\begin{document}

\title{High-statistics measurement of the $\eta\to 3\pi^0$ decay at the Mainz Microtron}

\author{S.~Prakhov}\thanks{Electronic address: prakhov@ucla.edu}\Mainz \\ \UCLA \\
\author{S.~Abt}\Basel \\
\author{P.~Achenbach}\Mainz \\
\author{P.~Adlarson}\Mainz \\
\author{F.~Afzal}\Bonn \\
\author{P.~Aguar-Bartolom\'e}\Mainz \\
\author{Z.~Ahmed}\Regina \\
\author{J.~Ahrens}\Mainz \\
\author{J.~R.~M.~Annand}\Glasgow \\
\author{H.~J.~Arends}\Mainz \\
\author{K.~Bantawa}\Kent \\
\author{M.~Bashkanov}\Edinburgh \\
\author{R.~Beck}\Bonn \\
\author{M.~Biroth}\Mainz \\
\author{N.~S.~Borisov}\Dubna \\
\author{A.~Braghieri}\Pavia \\
\author{W.~J.~Briscoe}\GWU \\
\author{S.~Cherepnya}\LPI \\
\author{F.~Cividini}\Mainz \\
\author{C.~Collicott}\Dalhousie \\ \Halifax \\
\author{S.~Costanza}\Pavia \\ \UniPavia \\
\author{A.~Denig}\Mainz \\
\author{M.~Dieterle}\Basel \\
\author{E.~J.~Downie}\GWU \\
\author{P.~Drexler}\Mainz \\
\author{M.~I.~Ferretti Bondy}\Mainz \\
\author{L.~V.~Fil'kov}\LPI \\
\author{A.~Fix}\Tomsk \\
\author{S.~Gardner}\Glasgow \\
\author{S.~Garni}\Basel \\
\author{D.~I.~Glazier}\Glasgow \\ \Edinburgh \\
\author{I.~Gorodnov}\Dubna \\
\author{W.~Gradl}\Mainz \\
\author{G.~M.~Gurevich}\INR \\
\author{C.~B.~Hamill}\Edinburgh \\
\author{L.~Heijkenskj\"old}\Mainz \\
\author{D.~Hornidge}\Sackville \\
\author{G.~M.~Huber}\Regina \\
\author{A.~K\"aser}\Basel\\
\author{V.~L.~Kashevarov}\Mainz \\ \Dubna \\
\author{S.~Kay}\Edinburgh \\
\author{I.~Keshelashvili}\Basel\\
\author{R.~Kondratiev}\INR \\
\author{M.~Korolija}\Zagreb \\
\author{B.~Krusche}\Basel \\
\author{A.~Lazarev}\Dubna \\
\author{V.~Lisin}\INR \\
\author{K.~Livingston}\Glasgow \\
\author{S.~Lutterer}\Basel \\
\author{I.~J.~D.~MacGregor}\Glasgow \\
\author{D.~M.~Manley}\Kent \\ 
\author{P.~P.~Martel}\Mainz \\ \Sackville \\
\author{J.~C.~McGeorge}\Glasgow \\
\author{D.~G.~Middleton}\Mainz \\ \Sackville \\
\author{R.~Miskimen}\Amherst \\
\author{E.~Mornacchi}\Mainz \\
\author{A.~Mushkarenkov}\Pavia \\ \Amherst \\ 
\author{A.~Neganov}\Dubna \\
\author{A.~Neiser}\Mainz \\ 
\author{M.~Oberle}\Basel \\
\author{M.~Ostrick}\Mainz \\  
\author{P.~B.~Otte}\Mainz \\
\author{D.~Paudyal}\Regina \\
\author{P.~Pedroni}\Pavia \\
\author{A.~Polonski}\INR \\  
\author{G.~Ron}\Jerusalem \\
\author{T.~Rostomyan}\Basel \\
\author{A.~Sarty}\Halifax \\
\author{C.~Sfienti}\Mainz \\
\author{V.~Sokhoyan}\Mainz \\
\author{K.~Spieker}\Bonn \\
\author{O.~Steffen}\Mainz \\
\author{I.~I.~Strakovsky}\GWU \\
\author{B.~Strandberg}\Glasgow \\
\author{Th.~Strub}\Basel \\
\author{I.~Supek}\Zagreb \\
\author{A.~Thiel}\Bonn \\
\author{M.~Thiel}\Mainz \\
\author{A.~Thomas}\Mainz \\   
\author{M.~Unverzagt}\Mainz \\ 
\author{Yu.~A.~Usov}\Dubna \\
\author{S.~Wagner}\Mainz \\
\author{N.~K.~Walford}\Basel \\ 
\author{D.~P.~Watts}\Edinburgh \\
\author{D.~Werthm\"uller}\Glasgow \\ \Basel \\
\author{J.~Wettig}\Mainz \\
\author{L.~Witthauer}\Basel \\ 
\author{M.~Wolfes}\Mainz \\
\author{L.~A.~Zana}\Edinburgh \\

\collaboration{A2 Collaboration at MAMI}

\date{\today}
         
\begin{abstract}
 The largest, at the moment, statistics of $7\times 10^6$ $\eta\to 3\pi^0$ decays,
 based on $6.2 \times 10^7$ $\eta$ mesons produced in the $\gamma p\to \eta p$ reaction,
 has been accumulated by the A2 Collaboration at the Mainz Microtron, MAMI.
 It allowed a detailed study of the $\eta\to 3\pi^0$ dynamics beyond
 its conventional parametrization with just the quadratic slope parameter
 $\alpha$ and enabled, for the first time, a measurement of the second-order term
 and a better understanding of the cusp structure in the neutral decay.
 The present data are also compared to recent theoretical calculations that
 predict a nonlinear dependence along the quadratic distance from
 the Dalitz-plot center.   
\end{abstract}

\maketitle

\section{Introduction}

 For decades, the $\eta\to 3\pi$ decay has attracted much attention from
 theoretical and experimental studies as it gives access to fundamental
 physical constants. This decay, which is forbidden by isospin symmetry,
 mostly occurs due to the difference in the mass of the $u$ and $d$ quarks,
 with $\Gamma(\eta \to 3\pi) \sim (m_d-m_u)^2$~\cite{Gasser}. 
 Therefore, a precision measurement of this decay can be used as a
 sensitive test for the magnitude of isospin breaking in the Quantum Chromodynamics
 (QCD) part of the Standard Model (SM) Lagrangian.
 At the same time, the actual $\eta\to 3\pi$ dynamics involve a strong impact
 from $\pi\pi$ final-state interactions, and the $m_d-m_u$ magnitude
 cannot be approached without a precise experimental measurement
 of the $\eta\to 3\pi$ Dalitz plots, the density of which provides
 the information needed. Theoretical calculations of strong-interaction processes
 at low energy, which could typically be performed by using
 Chiral Perturbation Theory ($\chi$PTh)~\cite{Gasser,Bijnens_2002,Bijnens_2007}, were
 not very successful at describing the $\eta\to 3\pi$ density distributions observed
 experimentally. The main reason was in the final-state rescattering effects,
 the calculation of which turned out to be more reliable
 with dispersion relations~\cite{Kambor,anisovich}, but still insufficient
 to describe the experimental data.
  Meanwhile, the experimental progress in both
  the precise determination of the $\pi\pi$ phase
 shifts~\cite{Colangelo_2001,Kaminski_2008,Garcia_2011} and high-statistics data
 on the $\eta\to 3\pi^0$ and $\eta\to \pi^+\pi^-\pi^0$
 decays~\cite{eta_slope_bnl,Unverzagt_2009,Prakhov_2009,Ambrosino_2008,Adlarson_2014,Anastasi_2016}
 renewed the interest in theoretical studies of the $\eta\to 3\pi$ decay
 ~\cite{Rusetsky_2009,Schneider_2011,Kampf_2011,Guo_2015,Guo_2017,Kolesar_2017,Colangelo_2017,Albaladejo_2017},
 which also included the extraction of the quark-mass ratio,
 $Q^2=(m^2_s - \bar{m}^2_{ud})/(m^2_d - m^2_u)$ with $\bar{m}_{ud}=(m_u + m_d)/2$, from the data.
   
 The function describing the density of the $\eta\to 3\pi$ Dalitz plot follows
 the standard parametrization for three-body decay, which 
 is a polynomial expansion of $|A(s_1,s_2,s_3)|^2$ around the center of the Dalitz plot,
 where $s_i=(P_\eta-p_i)^2$, with $p_i^2=M_i^2$.
 The parameters are usually normalized to be dimensionless.
 The standard variables introduced for the $\eta\to 3\pi$ decay are then
 $X=\sqrt{3}(T_1-T_2)/Q_\eta=\sqrt{3}(s_2 -s_1)/(2m_\eta Q_\eta)$ and
 $Y=3T_3/Q_\eta-1=3((m_\eta -m_{\pi^0})^2 -s_3)/(2m_\eta Q_\eta)-1$, 
 where $T_i$ is the kinetic energy of pion $i$ in the $\eta$ rest frame,
 and $Q_\eta=m_\eta-3m_{\pi^0}$ for the neutral decay and $Q_\eta=m_\eta-2m_{\pi^{\pm}}-m_{\pi^0}$
 for the charged decay. In addition, another dimensionless variable
 $z=X^2+Y^2=6\sum_{i=1}^3 (T_i +m_{\pi^0} -m_{\eta}/3)^2/Q_\eta^2=\rho^2/\rho^2_{\mathrm{max}}$
 was introduced to describe the $\eta\to 3\pi^0$ Dalitz-plot density in terms of
 the quadratic distance, $\rho^2$, from the plot center.
 For the neutral $\eta$ decay, its polynomial expansion
\begin{eqnarray}   
 & A(s_1,s_2,s_3) \sim 1 & +\alpha'\sum_{i=1}^3(s_i-s_0)^2  +\beta'\sum_{i=1}^3(s_i-s_0)^3 \nonumber \\
 & &  +\gamma'\sum_{i=1}^3(s_i-s_0)^4 +...~, \label{eg:as123}
\end{eqnarray}
 with $s_0 = m_\eta /3 -m_{\pi^0}$, results in~\cite{Schneider_2011,Kampf_2011}
\begin{equation}   
 |A(X,Y)|^2 \sim 1+ 2\alpha z + 2\beta (3X^2 Y-Y^3) + 2\gamma z^2 +...~,
\label{eg:axy}
\end{equation}
 where parameters $\alpha'$, $\beta'$, and $\gamma'$ are complex in general, and
 parameters $\alpha$, $\beta$, and $\gamma$ are real.
 Representing $X=\sqrt{z}\cos(\phi)$ and $Y=\sqrt{z}\sin(\phi)$ as polar coordinates
 with respect to the Dalitz-plot center, Eq.~(\ref{eg:axy}) can be rewritten as
\begin{equation}   
 |A(X,Y)|^2 \sim 1+ 2\alpha z + 2\beta z^{3/2} \sin(3\phi) + 2\gamma z^2 +...~,
\label{eg:azphi}
\end{equation}
 where angle $\phi = \arctan(Y/X)$.

\begin{figure}
\includegraphics[width=0.48\textwidth]{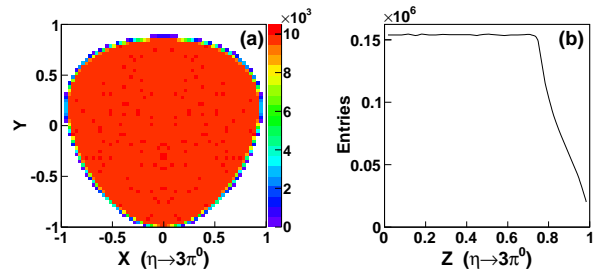}
\caption{
 Dalitz plot and its $z=X^2+Y^2=\rho^2/\rho^2_{\mathrm{max}}$
 distribution for the phase-space $\eta\to 3\pi^0$ decay. 
}
 \label{fig:dk_eta3pi0_dalpl_zeta} 
\end{figure}

 Due to the low energies of the decay pions, $\pi^0\pi^0$ rescattering in
 $\eta\to 3\pi^0$ is expected to be dominated by S waves.
 Such an assumption leads to the conventional leading-order parametrization
 $|A(z)|^2 \sim 1+ 2\alpha z$~\cite{PDG} of the $\eta\to 3\pi^0$ amplitude,
 with only the quadratic slope parameter $\alpha$, which was used in all previous measurements.
 Rather than fitting two-dimensional Dalitz plots, those measurements were based on
 the deviation of measured $z$ distributions from the corresponding
 distributions obtained from the phase-space simulation of the $\eta\to 3\pi^0$ decay,
 which is illustrated for both the Dalitz plot and $z$ distribution in
 Fig.~\ref{fig:dk_eta3pi0_dalpl_zeta}.

\begin{figure*}
\includegraphics[width=0.9\textwidth]{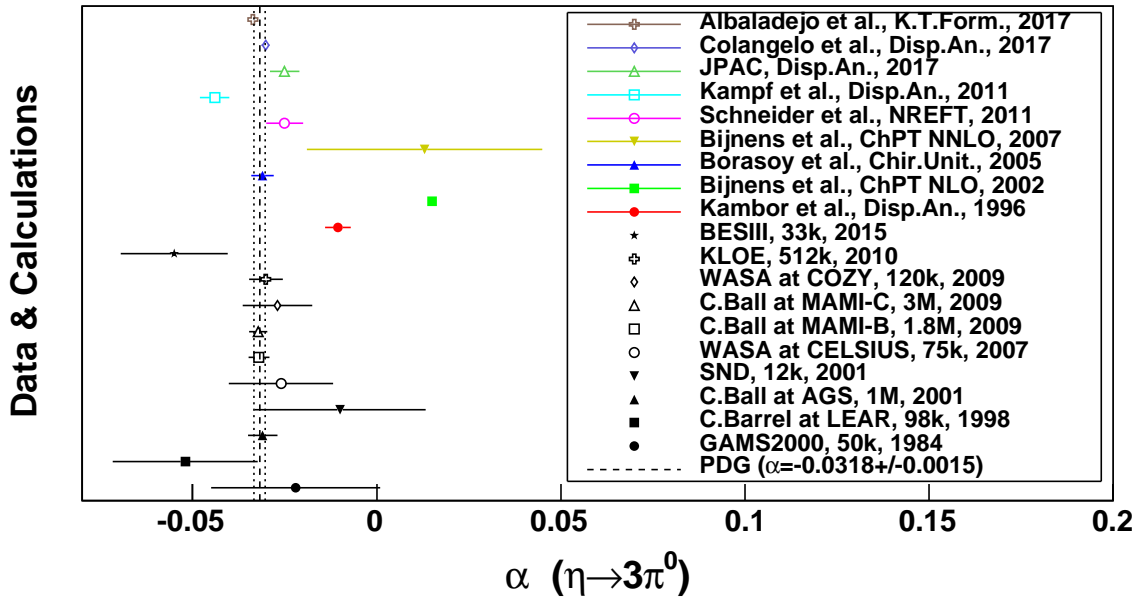}
\caption{ 
 Comparison of the experimental data~\protect\cite{eta_slope_bnl,Unverzagt_2009,Prakhov_2009,GAMS_1984,CBarrel_1998,SND_2001,WASA_2007,WASA_2009,KLOE_2011,BESIII_2015} (plotted by black points),
 which were used in the RPP~\protect\cite{PDG} to obtain the averaged value
 (shown by the vertical lines)
 for the $\eta\to 3\pi^0$ quadratic slope parameter $\alpha$, to each other and to various
 calculations~\protect\cite{Bijnens_2002,Bijnens_2007,Schneider_2011,Kampf_2011,Guo_2017,Colangelo_2017,Albaladejo_2017,Kambor,Borasoy}
 (colored points). Ref.~\protect\cite{Bijnens_2002} gives the magnitude of $\alpha$
 for the analysis made in Ref.~\protect\cite{Gasser}, in which its value
 was not given.
}
 \label{fig:alpha_eta3pi0_comp} 
\end{figure*}

 The current value for the $\eta\to 3\pi^0$ quadratic slope parameter,
 $\alpha=-0.0318\pm 0.0015$, which is given in the Review of Particle Physics (RPP)~\cite{PDG},
 is based on ten measurements~\cite{eta_slope_bnl,Unverzagt_2009,Prakhov_2009,GAMS_1984,CBarrel_1998,SND_2001,WASA_2007,WASA_2009,KLOE_2011,BESIII_2015}.
 The results of those measurements are plotted in Fig.~\ref{fig:alpha_eta3pi0_comp}
 along with values from various
 calculations~\cite{Bijnens_2002,Bijnens_2007,Schneider_2011,Kampf_2011,Guo_2015,Guo_2017,Colangelo_2017,Kambor,Borasoy}.
 As shown in Fig.~\ref{fig:alpha_eta3pi0_comp}, all experimental results obtained with
 comparably large statistics are in good agreement within their uncertainties,   
 and the earlier theoretical calculations contradict experimental data more than
 the most recent.

 The result with the best accuracy,
($\alpha=-0.0322\pm0.0012_{\mathrm{stat}}\pm0.0022_{\mathrm{syst}}$,
 obtained by the A2 Collaboration at MAMI, was based on $3\times 10^6$ observed
 $\eta\to 3\pi^0$ decays~\cite{Prakhov_2009}.
 Significant attention in that work was dedicated to a search for a possible
 cusp structure in the spectra below the $\pi^+\pi^-$ threshold.
 Based on the $\pi\pi$ scattering length combination $a_0 -a_2$,
 extracted from the analysis of $K \to 3\pi$ decays~\cite{NA48_2009},
 and calculations within the framework of nonrelativistic effective field
 theory (NREFT)~\cite{Rusetsky_2009}, the cusp effect was expected to be visible in
 the $m(\pi^0\pi^0)$ spectrum, reaching $\sim 1\%$ at the $2\pi^0$ threshold with respect
 to the spectrum in the case of the cusp absence.
 This calculation used the $\eta\to \pi^+\pi^-\pi^0$ results from KLOE~\cite{Ambrosino_2008}
 to describe the charge-decay amplitude, assuming the isospin limit to connect it
 to the neutral decay. In principle, the predicted cusp magnitude should not change
 much even in the case of isospin breaking.
 However, the expected cusp structure was not confirmed experimentally
 in Ref.~\cite{Prakhov_2009}. At the same time, the statistical accuracy of
 data points in the measured $z$ distribution made it possible to indicate
 that the conventional leading-order parametrization
 $|A(z)|^2 \sim 1+ 2\alpha z$ was not sufficient for the proper description of
 the $\eta\to 3\pi^0$ decay amplitude. This indicates that the contributions from
 the higher-order terms in Eq.~(\ref{eg:azphi}) need to be
 checked as well. The cusp structure cannot be described by polynomial expansion but,
 similar to the NREFT, the cusp range can be parametrized in the density function as
 $\rho(s) = {\rm Re}\sqrt{(1-s/4m^{2}_{\pi^{\pm}})}$,
 which results in $\rho(s) = 0$ for $s\ge 4m^{2}_{\pi^{\pm}}$~\cite{Heinrich_priv}.
 Then the density function is given by
 \begin{eqnarray}   
 & |A|^2 \sim 1 & + 2\alpha z + 2\beta z^{3/2} \sin(3\phi) + 2\gamma z^2 +... \nonumber \\
 &            &   + 2\delta\sum_{i=1}^3\rho(s_i)~,
\label{eg:azphicu}
\end{eqnarray}
 where the factor 2 in front of the cusp term is added for the consistency
 with the other terms.

 A better determination of the $\eta \to 3\pi$ decay parameters, needed for
 a precise determination of light-quark mass ratios, was recently the focus of
 many theoretical works. In Ref.~\cite{Schneider_2011}, a detailed study of
 the $\eta \to 3\pi$ decays within the framework of the modified NREFT,
 in which final-state interactions were analyzed beyond one loop including
 isospin-breaking corrections, resulted in the extraction of the Dalitz-plot
 parameters for both the charged and neutral decays.
 The values obtained for the parametrization of the neutral decay with
 Eq.~(\ref{eg:azphi}), $\alpha=-0.0246(49)$, $\beta=-0.0042(7)$, and $\gamma=0.0013(4)$,
 indicated nonzero contributions for the higher-order terms. 
 Other $\eta\to 3\pi^0$ calculations, involving parameter $\beta$, used
 a unitary dispersive model~\cite{Guo_2015,Guo_2017}, in which
 substraction constants were fixed by fitting recent high-statistics
 $\eta\to \pi^+\pi^-\pi^0$ data from WASA-at-COSY
 ($1.74\times 10^5$ decays)~\cite{Adlarson_2014} and
 KLOE ($4.7\times 10^6$ decays)~\cite{Anastasi_2016}.
 In contrast to Ref.~\cite{Schneider_2011}, the latter calculations
 predicted a value of $\beta$ consistent with zero.
 Another recent dispersive analysis~\cite{Colangelo_2017} of the $\eta\to 3\pi$
 decay amplitudes, in which the latest $\eta\to \pi^+\pi^-\pi^0$ data
 from KLOE~\cite{Anastasi_2016} were also fitted to determine subtraction constants,
 predicted a nonlinear $z$ dependence for $\eta\to 3\pi^0$, which turned out to be
 in good agreement within the uncertainties with the measured $z$ dependence 
 from Ref.~\cite{Prakhov_2009}. However, no numerical predictions were provided
 for the higher-order terms of Eq.~(\ref{eg:azphi}).
 The most recent $\eta\to 3\pi$ calculation, which used the extended chiral
 Khuri-Treiman dispersive formalism~\cite{Albaladejo_2017}, showed that
 the effect from the two light resonances $f_0(980)$ and $a_0(980)$
 in the low energy region of the $\eta\to 3\pi$ decay is not negligible,
 especially for the neutral mode, and improves the description of
 the density variation over the Dalitz plot. 
 The $\eta\to 3\pi^0$ parameters obtained in Ref.~\cite{Albaladejo_2017}
 from their fitted amplitude, $\alpha=-0.0337(12)$ and $\beta=-0.0054(1)$,
 also predict nonzero contributions for the $2\beta z^{3/2} \sin(3\phi)$ term. 

 Obviously, a better comparison of the experimental data with
 the recent $\eta\to 3\pi^0$ calculations, going beyond
 the leading-order parametrization, should now be based on
 describing the two-dimensional density distribution
 of measured Dalitz plots, rather than on one-dimensional $z$ distributions.
 To obtain reliable experimental results for the parametrization
 with Eq.~(\ref{eg:azphicu}),  a new measurement of the $\eta\to 3\pi^0$ Dalitz
 plot, with even higher statistical accuracy, is very important.
 
 In this paper, we report on a new high-statistics measurement
 of the $\eta \to 3\pi^0$ Dalitz plot, which is based on
 $7\times 10^6$ detected decays. The A2 data used in the present analysis
 were taken in 2007 (Run I) and 2009 (Run II).
 Compared to the previous analysis of Run I reported in Ref.~\cite{Prakhov_2009},
 the present analysis was made with an improved cluster algorithm, which
 increased the number of $\eta \to 3\pi^0$ decays reconstructed in Run I from $3\times 10^6$
 to $3.5\times 10^6$. The $\gamma p \to \eta p \to 3\pi^0 p \to 6\gamma p$ data
 from Run I and Run II used in this work were previously used to measure
 the $\gamma p \to \eta p$ differential cross sections, the analysis of which
 was recently reported in Ref.~\cite{A2_eta_etapr_2017}.
 The new $\eta \to 3\pi^0$ results were obtained
 with the parametrization involving the higher-order terms of the Dalitz-plot
 density function and the cusp term.
 The NREFT framework from Ref.~\cite{Bissegger} was also used to check
 whether the present $\eta\to 3\pi^0$ data can be described together
 with the KLOE $\eta\to \pi^+\pi^-\pi^0$ data~\cite{Anastasi_2016}, assuming the isospin limit.
 The experimental spectra are also compared to recent theoretical calculations that
 predict a nonlinear dependence along the quadratic distance from
 the Dalitz-plot center.   
   
\section{Experimental setup}
\label{sec:Setup}

 An experimental study of the $\eta \to 3\pi^0$ decay was conducted via
 measuring the process $\gamma p \to \eta p \to 3\pi^0 p \to 6\gamma p$
 with the Crystal Ball (CB)~\cite{CB} as a central calorimeter and
 TAPS~\cite{TAPS,TAPS2} as a forward calorimeter. These detectors were
 installed in the energy-tagged bremsstrahlung photon beam of
 the Mainz Microtron (MAMI)~\cite{MAMI,MAMIC}. 
 The photon energies were determined
 by the Glasgow tagging spectrometer~\cite{TAGGER,TAGGER1,TAGGER2}.

 The CB detector is a sphere consisting of 672
 optically isolated NaI(Tl) crystals, shaped as
 truncated triangular pyramids, which point toward
 the center of the sphere. The crystals are arranged in two
 hemispheres that cover 93\% of $4\pi$, sitting
 outside a central spherical cavity with a radius of
 25~cm, which holds the target and inner
 detectors. In this experiment, TAPS was initially
 arranged in a plane consisting of 384 BaF$_2$
 counters of hexagonal cross section.
 It was installed 1.5~m downstream of the CB center
 and covered the full azimuthal range for polar angles
 from $1^\circ$ to $20^\circ$. 
 Later on, 18 BaF$_2$ crystals,
 covering polar angles from $1^\circ$ to $5^\circ$, were replaced
 with 72 PbWO$_4$ crystals, allowing for a higher count rate
 in the crystals near the photon-beam line.
 More details on the energy and angular resolution of the CB and TAPS
 are given in Refs.~\cite{Prakhov_2009,etamamic}.

 The present measurement used electron beams
 with energies of 1508 and 1557 MeV from the Mainz Microtron, MAMI-C~\cite{MAMIC}.
 The data with the 1508-MeV beam were taken in 2007 (Run I)
 and those with the 1557-MeV beam in 2009 (Run II).
 Bremsstrahlung photons, produced by the beam electrons
 in a 10-$\mu$m Cu radiator and collimated by a 4-mm-diameter Pb collimator,
 were incident on a liquid hydrogen (LH$_2$) target located
 in the center of the CB. The LH$_2$ target was 5 cm and 10 cm long
 in Run I and Run II, respectively.
 The total amount of material around the LH$_2$ target,
 including the Kapton cell and the 1-mm-thick carbon-fiber beamline,
 was equivalent to 0.8\% of a radiation length $X_0$,
 which was essential to keep the material budget
 as low as possible to minimize the conversion of final-state photons.

 The target was surrounded by a Particle IDentification
 (PID) detector~\cite{PID} used to distinguish between charged and
 neutral particles. The PID consists of 24 scintillator bars
 (50 cm long, 4 mm thick) arranged as a cylinder with the middle radius of 12 cm.

 In Run I, the energies of the incident photons were analyzed
 up to 1402~MeV by detecting the postbremsstrahlung electrons
 in the Glasgow tagged-photon spectrometer
 (Glasgow tagger)~\cite{TAGGER,TAGGER1,TAGGER2},
 and up to 1448~MeV in Run II.
 The uncertainty in the energy of the tagged photons is mainly determined
 by the segmentation of the tagger focal-plane detector in combination with
 the energy of the MAMI electron beam used in the experiments.
 Increasing the MAMI energy increases the energy range covered
 by the spectrometer and also has the corresponding effect on the uncertainty
 in $E_\gamma$. For both the MAMI energy settings of 1508 and 1557~MeV,
 this uncertainty was about $\pm 2$~MeV.
 More details on the tagger energy calibration and uncertainties
 in the energies can be found in Ref.~\cite{TAGGER2}.

 The experimental trigger in Run I required the total energy deposited in the CB
 to exceed $\sim$320~MeV and the number of so-called hardware clusters
 in the CB (multiplicity trigger) to be two or more.
 In the trigger, a hardware cluster in the CB was a block of 16
 adjacent crystals in which at least one crystal had an energy
 deposit larger than 30 MeV.
 Depending on the data-taking period, events with a cluster multiplicity
 of two were prescaled with different rates.
 TAPS was not included in the multiplicity trigger for these experiments.
 In Run II, the trigger threshold on the total energy
 in the CB was increased to $\sim$340~MeV, and the multiplicity
 trigger required three or more hardware clusters in the CB.

\section{Data analysis}

 The $\eta \to 3\pi^0$ decays were measured via the process
 $\gamma p \to \eta p\to 3\pi^0 p \to 6\gamma p$
 from events having six or seven clusters reconstructed by a software analysis
 in the CB and TAPS together. Seven-cluster events were analyzed by assuming that all
 final-state particles were detected, and six-cluster events by assuming that
 only the six photons were detected, with the recoil proton going undetected. 
 The offline cluster algorithm~\cite{k0sn_lpi0_spi0_2009} was optimized for finding
 a group of adjacent crystals in which the energy was deposited
 by a single-photon electromagnetic (e/m) shower.
 This algorithm also works well for recoil protons.
 The software threshold for the cluster energy was chosen to be 12 MeV.
 Compared to the previous $\eta \to 3\pi^0$ analysis of Run I~\cite{Prakhov_2009},
 the cluster algorithm was improved for a better separation
 of e/m showers partially overlapping in the calorimeters,
 which is especially important for processes with large
 photon multiplicity in the final state and for conditions of
 the forward energy boost of the outgoing photons
 in the laboratory system. At the same time, the cluster algorithm has also to be
 efficient for reconstructing one photon splitting into two nearby e/m showers.
 The new optimization of the cluster algorithm was needed to improve its efficiency
 for higher energies of MAMI-C.
 Particularly for the process $\gamma p \to \eta p \to 3\pi^0 p \to 6\gamma p$,
 its reconstruction efficiency was improved by $\sim 17\%$, compared
 to the previous analysis~\cite{Prakhov_2009}.
   
 The event identification was based on a kinematic fit, the details of which,
 including the parametrization of the detector
 information and resolutions were given in Ref.~\cite{Prakhov_2009}.
 Many other details of the event selection in the present work are also very similar
 to the previous analysis.
 To test the $\gamma p \to \eta p \to 3\pi^0 p \to 6\gamma p$ hypothesis,
 15 combinations are possible to pair six photons into three neutral pions.
 To reduce the number of combinations tested with the kinematic fit,
 invariant masses of cluster pairs for each combination were tested prior to fitting.
 For seven-cluster events, where seven combinations are possible to select the proton cluster,
 this number was reduced by a cut on the cluster polar angle,
 the value of which is limited by the recoil-proton kinematics in the laboratory system. 
 The events for which at least one pairing combination satisfied
 the tested hypothesis at the 1\% confidence level, CL, (i.e., with a probability greater
 than 1\%) were selected for further analysis. The pairing combination with
 the largest CL was used to reconstruct the reaction kinematics.
 The combinatorial background from mispairing six photons into three pions was found
 to be quite small and could be further reduced by tightening a selection criterion on
 the kinematic-fit CL. Misidentification of the proton cluster with the photons
 was found to be negligibly small for seven-cluster events. The six-cluster sample,
 which includes $\sim 20\%$ from all detected $\eta \to 3\pi^0$ decays,
 had a small contamination from events in which one of the photons, instead of
 the proton, was undetected. Because such misidentification mostly occurred for
 clusters in TAPS, those events were successfully removed, based on the cluster's
 time-of-flight information, which provides good separation of
 the $\gamma p \to \eta p$ recoil protons from photons in the present energy range. 

 To minimize systematic uncertainties in the determination of
 experimental acceptance, Monte Carlo (MC) simulations of the production reaction
 $\gamma p \to \eta p$ were based on the actual spectra measured with the same
 data sets~\cite{A2_eta_etapr_2017}. The $\eta \to 3\pi^0$ decay was generated according
 to phase space (i.e., with the slope parameter $\alpha=0$). The simulated events
 were propagated through a {\sc GEANT} (version 3.21) simulation of the experimental
 setup. To reproduce the resolutions observed in the experimental data, the {\sc GEANT}
 output (energy and timing) was subject to additional smearing, thus
 allowing both the simulated and experimental data to be analyzed in the same way.
 Matching the energy resolution between the experimental and MC events
 was achieved by adjusting the invariant-mass resolutions,
 the kinematic-fit stretch functions (or pulls), and probability
 distributions. Such an adjustment was based on the analysis of the
 same data sets for reactions that could be selected with the kinematic fit
 practically without background from other reactions
 (namely, $\gamma p\to \pi^0 p$, $\gamma p\to \eta p\to \gamma\gamma p$,
 and $\gamma p\to \eta p\to 3\pi^0p$ were used).
 The simulated events were also tested to check whether they passed
 the trigger requirements.
 
\begin{figure}
\includegraphics[width=0.48\textwidth]{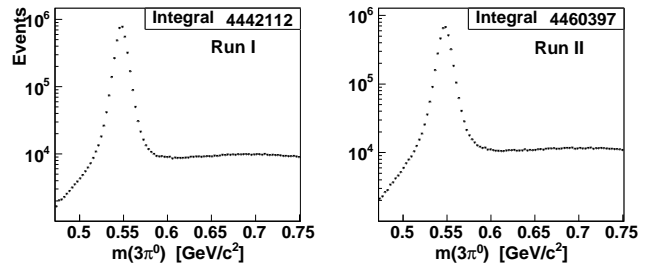}
\caption{
 $m(3\pi^0)$ invariant-mass distributions for events selected at CL$>1\%$
 by testing the $\gamma p \to 3\pi^0 p \to 6\gamma p$ hypothesis for the
 data of Run I (left) and Run II (right).
}
 \label{fig:m3pi0_2007_09} 
\end{figure}

 For $\eta\to 3\pi^0$ decays, physical background can only come from
 the $\gamma p\to 3\pi^0p$ events that are not produced from $\eta$ decays.
 As shown in Ref.~\cite{K0Sigpl2013}, those $3\pi^0$ events are mostly produced
 via baryon decay chains, with a smaller fraction from
 $\gamma p\to K^0_S \Sigma^+ \to 3\pi^0p$. For selected $\gamma p \to \eta p \to 3\pi^0 p$
 events, this background is negligibly small near the $\eta$ production threshold,
 and reaches $\sim 4\%$ near beam energy $E_\gamma=1.4$~GeV.
 Because of the complicated dynamics of these background processes, they cannot be reproduced
 precisely with the MC simulation in order to be used for the background subtraction, and
 additional selection criteria have to be applied instead to reduce the remaining background
 to a level $\le 1\%$.
 The initial level of the direct $3\pi^0$ background under the $\eta\to 3\pi^0$ peak
 can be seen in the $m(3\pi^0)$ invariant-mass distributions for events selected at CL$>1\%$
 by testing the $\gamma p \to 3\pi^0 p \to 6\gamma p$ hypothesis, which has
 no constraint on the $\eta$ mass. These distributions are shown in
  Fig.~\ref{fig:m3pi0_2007_09}.
 It was checked that the level of the direct $3\pi^0$ background $\le 1\%$ in the
 $\eta\to 3\pi^0$ data sample could be reached by requiring CL$>1.5\%$ for
 the $\gamma p \to \eta p \to 3\pi^0 p \to 6\gamma p$ hypothesis along with rejecting
 events having $E_\gamma > 1.3$~GeV. 
  
 There are two more sources of background remaining in the selected
 $\gamma p \to \eta p\to 3\pi^0 p \to 6\gamma p$ events and which could directly be
 subtracted from the experimental spectra.
 The first background is due to interactions of the bremsstrahlung photons
 in the windows of the target cell.
 The evaluation of this background is based on the
 analysis of data samples that were taken with the target cell emptied of
 liquid hydrogen. The weight for the subtraction
 of empty-target spectra is usually taken as a ratio of the photon-beam
 fluxes for the data samples with the full and the empty target.
 Because, in the present experiments, the amount of empty-target data were
 much smaller than with the full target, the subtraction of this background
 would cause larger statistical uncertainties.
 It was checked that, for the selection criteria used,
 the fraction of the empty-target background is $\le 1\%$, and this background
 mostly contains actual $\eta\to 3\pi^0$ decays that were just produced
 in interactions with the target-cell material. Thus, the subtraction
 of the empty-target background was neglected in the present analysis.  
\begin{figure*}
\includegraphics[width=0.98\textwidth]{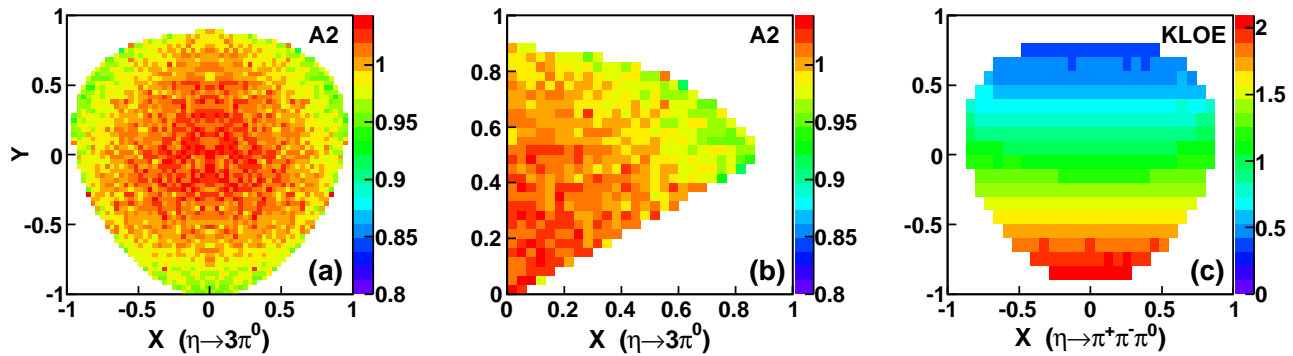}
\caption{ 
 Comparison of the experimental $\eta\to 3\pi$ Dalitz plots:
 (a) the full $\eta\to 3\pi^0$ plot (six entries per decay) from the present analysis
  of $\sim 7\times10^6$ decays;
 (b) one sextant of the $\eta\to 3\pi^0$ plot (one entry per decay)
    for the angle range $30^\circ<\phi<90^\circ$ in Eq.~(\protect\ref{eg:azphi});
 (c) the $\eta\to \pi^+\pi^-\pi^0$ plot (without boundary bins) from
 the KLOE analysis of $\sim 4.7\times10^6$ decays~\protect\cite{Anastasi_2016}.
}
 \label{fig:accor_eta3pi_dalpl} 
\end{figure*}

 The second background was caused by random coincidences
 of the tagger counts with the experimental trigger.
 It mostly includes $\gamma p \to \eta p \to 3\pi^0 p \to 6\gamma p$ events
 reconstructed with random $E_\gamma$, resulting in poorer $\chi^2$ and resolution
 after kinematic fitting.  The subtraction of this background was carried out
 by using event samples for which all coincidences were random
 (see Ref.~\cite{Prakhov_2009} for more details).
 The fraction of random background was 6.7\% for Run I, and 6.9\% for Run II.
 The actual background samples included much more events to diminish
 the impact from statistical fluctuations in the distributions used
 for the subtraction.
 
\section{Results and Discussion}

 The full Dalitz plot obtained from $\sim 7\times10^6$ $\eta\to 3\pi^0$
 decays of Run I and Run II is shown in Fig.~\ref{fig:accor_eta3pi_dalpl}(a).
 Because there are three identical particles in the final state, variables $X$ and $Y$ can
 be determined in six different ways, with the same value for variable $z$ and
 different angle $\phi$ from Eq.~(\ref{eg:azphi}). Each of these six combinations
 in $X$ and $Y$ goes into six different sextants, repeating the density structure
 every 60 degrees. The difference between those sextants is only in their different
 orientation with respect to each other and to the plot binning.
 Also, this Dalitz plot is symmetric with respect to the Y axis.
 In principle, one sextant is sufficient to analyze the Dalitz-plot shape and
 to obtain the corresponding results with proper statistical uncertainties.
 Such a sextant plot, obtained for the angle range $30^\circ<\phi<90^\circ$,
 is shown in Fig.~\ref{fig:accor_eta3pi_dalpl}(b).
 As seen, this sextant plot has bins with limited physical coverage not
 only along the external edge but also along angle $\phi=30^\circ$.
 To avoid any dependence of the results on such an effect and on
 the sextant orientation with respect to the plot binning,
 one half of the Dalitz plot ($X<0$ or $X>0$) can be used to
 analyze its shape. Because half of the plot has three entries per event, the parameter
 errors from fitting to such a plot must be multiplied by the factor of $\sqrt{3}$
 to reflect the actual experimental statistics.
\begin{figure*}
\includegraphics[width=0.95\textwidth]{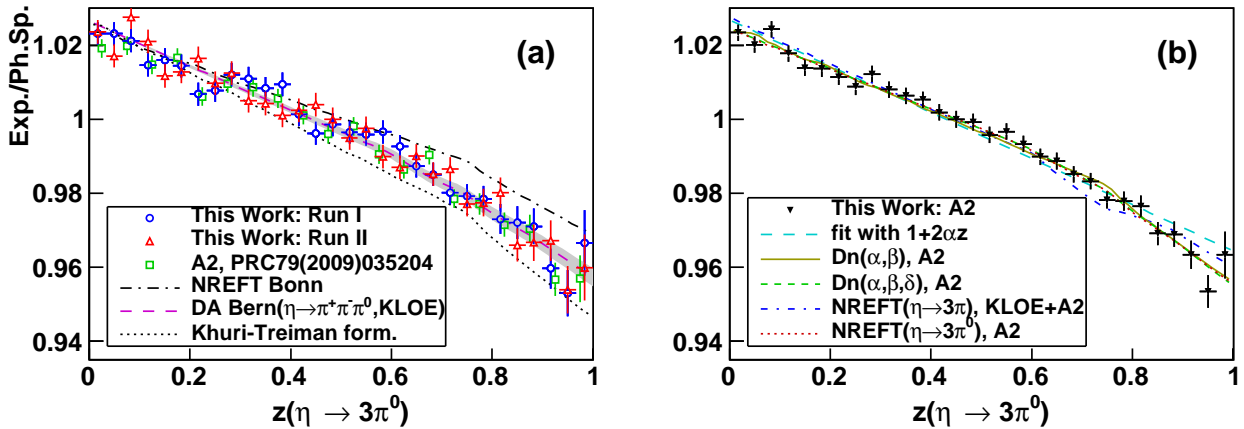}
\caption{ 
 Experimental $z$ distributions obtained (a) individually from Run I (blue circles)
 and Run II (red triangles), and (b) the combined results (black triangles).
 The earlier A2 data from Ref.~\protect\cite{Prakhov_2009} are depicted in (a)
 by green open squares. 
 The NREFT calculation by the Bonn group~\protect\cite{Schneider_2011} is shown in (a)
 by the black long-dash-dotted line.
 The prediction from the dispersive analysis by the Bern group~\protect\cite{Colangelo_2017}
 is shown in (a) by the magenta long-dashed line with an error band.
 The prediction based on the extended chiral Khuri-Treiman formalism~\protect\cite{Albaladejo_2017}
 is shown in (a) by the black dotted line.
 The fit of the combined $z$ distribution with the leading-order term
 (fit no.~2 in Table~\protect\ref{tab:rslt}) is shown in (b)
 by the cyan long-dashed line. The fits of the Dalitz-plot sextant with
 Eq.~(\protect\ref{eg:azphicu}), namely nos.~4 and 6 from Table~\protect\ref{tab:rslt},
 are shown in (b) by the yellow solid and the green dashed lines, respectively.
 The isospin-limit results from fitting both the present $\eta\to 3\pi^0$ and KLOE's
 $\eta\to \pi^+\pi^-\pi^0$~\protect\cite{Anastasi_2016} data within the NREFT framework
 from Ref.~\protect\cite{Bissegger} are shown by the blue dash-dotted line.
 The isospin-breaking results from fitting solely the $\eta\to 3\pi^0$ data within
 the same NREFT framework are shown by the red dotted line. 
}
 \label{fig:zeta_eta3pi0_a2_comp} 
\end{figure*}

 To obtain the $\eta\to 3\pi^0$ plots shown in Figs.~\ref{fig:accor_eta3pi_dalpl}(a)
 and \ref{fig:accor_eta3pi_dalpl}(b),
 the plots with the measured decays from Runs I and II were divided by the corresponding
 plots obtained from the analysis of the $\gamma p \to \eta p \to 3\pi^0 p$ MC simulations
 for those data sets. Because $\eta \to 3\pi^0$ decays were generated as phase space,
 the ratio of the experimental and the MC plots provides both the acceptance correction
 for the full area and the cancellation of the phase-space factor coming from
 the limited physical coverage, which is typical for boundary bins.
 Then those boundary bins can be treated in the same way as the inner bins
 while fitting the acceptance-corrected Dalitz plots with density functions.
 The only difference from the inner bins is
 in using $X$ and $Y$ coordinates averaged inside the boundary bins over the available
 phase space, instead of taking the bin centers.
 To combine the acceptance-corrected plots from different data sets
 (namely from Runs I and II), their normalization should be done in the same way.
 In the present analysis, an identical normalization was made by taking the 
 weight of the MC Dalitz plot as the ratio of the event numbers in the experimental
 and the MC plots.

 As shown in Fig.~\ref{fig:accor_eta3pi_dalpl}(a), the largest density of events
 is accumulated in the center of the $\eta\to 3\pi^0$ Dalitz plot, with a smooth
 decrease of a few percent toward the plot edge. To compare such a structure with
 the charged decay, the acceptance-corrected $\eta\to \pi^+\pi^-\pi^0$ Dalitz plot
 from KLOE~\cite{Anastasi_2016} (with excluded boundary bins)
 is illustrated in Fig.~\ref{fig:accor_eta3pi_dalpl}(c), showing
 a sharp decrease in its density from the smallest $Y$ to the largest.
 In the present work, this $\eta\to \pi^+\pi^-\pi^0$ plot was used to check whether
 it could be described together with the $\eta\to 3\pi^0$ data within the NREFT
 framework~\cite{Bissegger}, assuming the isospin limit.

 The advantage of analyzing the $\eta\to \pi^+\pi^-\pi^0$ decay is the fact that
 the $X$ and $Y$ variables can be defined uniquely. Then the experimental raw
 (i.e., uncorrected for the acceptance) Dalitz plot can be fitted with
 the corresponding plots of the phase-space MC events weighted with the density-function terms.
 Because the weights are calculated from the generated variables, but filling the MC plots is done
 according to the reconstructed variables, such a fit takes into account both
 the experimental acceptance and resolution. 
 For the $\eta\to 3\pi^0$ decay, the $X$ and $Y$ generated in one sextant could be reconstructed
 in another sextant, which allows proper fitting a sextant of the raw Dalitz plot
 with the density function dependent only on $z$ (which is the same
 for all pairs of $X$ and $Y$) but not on $\phi$.
 Therefore, all fits with the higher-order terms were made only for the
 acceptance-corrected Dalitz plots. The sensitivity of the results to the
 experimental resolution, which could be determined by comparing to the fits to
 the raw Dalitz plots, was only checked for the leading-order parametrization. 

 The traditional $z$ distributions, which were used in all previous measurements of
 the slope parameters $\alpha$, were obtained individually for Run I and Run II.
 Similar to the individual Dalitz plots, their normalization was based on the
 ratio of the total number of events in the experimental and the MC distributions,
 which allows the proper combination of the two independent measurements. 
 The individual $z$ distributions from Run I and Run II are compared
 in Fig.~\ref{fig:zeta_eta3pi0_a2_comp}(a)
 with each other and with the earlier A2 data from Ref.~\cite{Prakhov_2009},
 demonstrating good agreement within their statistical uncertainties.
 The combined $z$ distribution, shown
 in Fig.~\ref{fig:zeta_eta3pi0_a2_comp}(b), has a statistical
 accuracy in its 30 data points that appears to be sufficient to reveal the deviation
 from a linear dependence.  

\begin{figure*}
\includegraphics[width=0.95\textwidth]{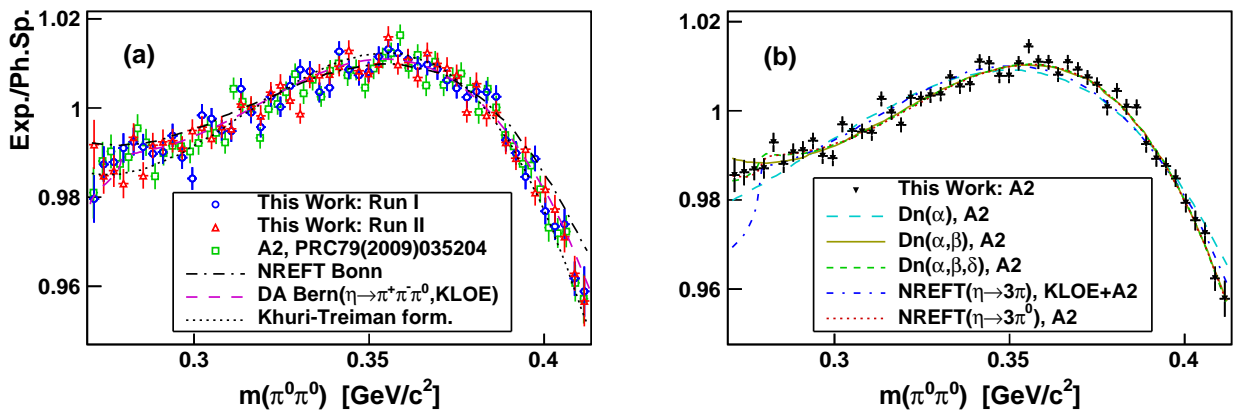}
\caption{ 
 Ratios of the experimental $m(\pi^0\pi^0)$ invariant-mass distributions to phase space
 obtained (a) individually from Run I (blue circles) and Run II (red triangles),
 and (b) the combined results (black triangles).
 The earlier A2 data from Ref.~\protect\cite{Prakhov_2009} are depicted in (a)
 by green open squares. 
 The NREFT calculation by the Bonn group~\protect\cite{Schneider_2011} is shown in (a)
 by the black long-dash-dotted line.
 The prediction from the dispersive analysis by the Bern
 group~\protect\cite{Colangelo_2017,Heinrich_priv}
 is shown in (a) by the magenta long-dashed line.
 The prediction based on the extended chiral Khuri-Treiman formalism~\protect\cite{Albaladejo_2017}
 is shown in (a) by the black dotted line.
 The combined $m(\pi^0\pi^0)$ distribution is compared in (b) to the results
 of fitting a sextant ($30^\circ<\phi<90^\circ$) of the acceptance-corrected $\eta\to 3\pi^0$
 Dalitz plot with the density function of Eq.~(\protect\ref{eg:azphicu}):
 fits no.~1 (cyan long-dashed line), no.~4 (yellow solid line), and no.~6 (green dashed line)
 in Table~\protect\ref{tab:rslt}.
 The isospin-limit results from fitting both the present $\eta\to 3\pi^0$ and KLOE's
 $\eta\to \pi^+\pi^-\pi^0$~\protect\cite{Anastasi_2016} data within the NREFT framework
 from Ref.~\protect\cite{Bissegger} are shown by the blue dash-dotted line.
 The isospin-breaking results from fitting solely the $\eta\to 3\pi^0$ data within
 the same NREFT framework are shown by the red dotted line. 
}
 \label{fig:m2pi0_eta3pi0_a2_comp} 
\end{figure*}

 The ratios of the experimental $m(\pi^0\pi^0)$ invariant-mass distributions to phase space,
 in which a cusp structure is expected to be seen, were obtained in the same way
 as the $z$ distributions. The agreement of the individual $m(\pi^0\pi^0)$ distributions
 from Run I, Run II, and the earlier A2 data from Ref.~\cite{Prakhov_2009},
 can be seen in Fig.~\ref{fig:m2pi0_eta3pi0_a2_comp}(a). 
 The combined $m(\pi^0\pi^0)$ distribution is shown in Fig.~\ref{fig:m2pi0_eta3pi0_a2_comp}(b),
 significantly improving the statistical accuracy in the cusp region, compared to the
 previous measurement~\cite{Prakhov_2009}.

 In addition to fitting the present $\eta\to 3\pi^0$ data with
 the density function from Eq.~(\protect\ref{eg:azphicu}), 
 the NREFT framework from Ref.~\cite{Bissegger} was used to check
 whether the neutral-decay data can be fitted well together
 with the KLOE $\eta\to \pi^+\pi^-\pi^0$ data~\cite{Anastasi_2016} by assuming the isospin limit.
 Next, the solely $\eta\to 3\pi^0$ data were fitted in the same framework by assuming 
 isospin breaking. In Ref.~\cite{Bissegger}, the decay amplitude is decomposed into up to
 two loops, $A(\eta\to 3\pi) = A^{\rm tree} + A^{\rm 1-loop} + A^{\rm 2-loop}$,
 with the tree amplitude complemented by final-state interactions of one
 and two loops. The tree amplitudes are parametrized as
 $A^{\rm tree}(\eta\to 3\pi^0) = K_0 + K_1(T_1^2 +T_2^2 +T_3^2)$ and
 $A^{\rm tree}(\eta\to \pi^+\pi^-\pi^0) = L_0 + L_1T_3 + L_2T_3^2 + L_3(T_1 -T_2)^2$,
 where $T_i = E_i - m_{\pi}$ is the kinetic energy of pion $i$ in the $\eta$ rest frame.
 For the conventional Dalitz plot variables, the tree amplitudes can be rewritten
 as $A^{\rm tree}(\eta\to 3\pi^0) = u_0 + u_1 z$ and
 $A^{\rm tree}(\eta\to \pi^+\pi^-\pi^0) = v_0 + v_1 Y + v_2 Y^2 + v_3 X^2$, where, at the tree level,
 the quadratic slope parameter is $\alpha = u_1 /u_0$, and the coefficients $u_i$ and $v_i$
 are strictly connected to $K_i$ and $L_i$, respectively. 
 Note that the shape of the actual $\eta\to 3\pi^0$ Dalitz plot is determined by
 the total amplitude; therefore, a measured $\alpha$ could be different from the
 ratio $u_1 /u_0$ of the tree-amplitude coefficients.
 The coefficients $K_i$ and $L_i$ (or $u_i$ and $v_i$ ) are also involved in the calculation
 of $A^{\rm 1-loop}$ and $A^{\rm 2-loop}$ for both the neutral and charged decays.
 The cusp structure below $2m_{\pi^{\pm}}$ appears
 in $A(\eta\to 3\pi^0)^{\rm 1-loop}$, and the cusp sign and magnitude is mostly determined by
 the scattering length combination $a_2 -a_0$~\cite{NA48_2009} and
 the $\eta\to \pi^+\pi^-\pi^0$ tree-amplitude coefficients $L_i$.
 In the isospin limit, the coefficients of the tree amplitude for the neutral decay
 can be rewritten via the coefficients of the charged decay:
 $K_0 = -(3L_0 +L_1 Q_\eta - L_3 Q_{\eta}^2)$ and $K_1 = -(L_2 +3L_3)$~\cite{Rusetsky_2009},
 with $Q_\eta=m_\eta-3m_{\pi^0}$.
 The isospin-limit fit to both the $\eta\to 3\pi^0$ and $\eta\to \pi^+\pi^-\pi^0$
 Dalitz plots has only five free parameters ($L_{i=1,2,3}$ and two normalization parameters),
 with fixed $L_0 =1$. The $\eta\to 3\pi^0$ data can also be fitted independently
 of the $\eta\to \pi^+\pi^-\pi^0$ decay by assuming isospin breaking,
 which requires the addition of $K_0$ and $K_1$ as free parameters,
 but leaves just one normalization parameter.

 Consistency of the present results for $z$ and $m(\pi^0\pi^0)$ with 
 theoretical calculations that predict a nonlinear $z$
 dependence~\cite{Schneider_2011,Colangelo_2017,Albaladejo_2017}
 is illustrated in Figs.~\ref{fig:zeta_eta3pi0_a2_comp}(a) and~\ref{fig:m2pi0_eta3pi0_a2_comp}(a). 
 The results of fits to the present data with various density functions,
 including the NREFT fits, are
 depicted in Figs.~\ref{fig:zeta_eta3pi0_a2_comp}(b) and~\ref{fig:m2pi0_eta3pi0_a2_comp}(b).
 The fit results with the density function from Eq.~(\protect\ref{eg:azphicu})
 are also listed in Table~\ref{tab:rslt} for different combinations of the density-function
 terms involved in a particular fit. 

 Fit no.~1 in Table~\ref{tab:rslt} was made to a sextant ($30^\circ<\phi<90^\circ$)
 of the acceptance-corrected Dalitz plot with the density function including
 only the leading-order term. Fit no.~2 was similar, but to
 the acceptance-corrected $z$ distribution as in all previous
 measurements. As shown, the values obtained there for $\alpha$ are practically
 the same and are in agreement within the fit errors with the RPP value
 $\alpha=-0.0318\pm 0.0015$~\cite{PDG}. The magnitudes of the fit $\chi^2$/ndf values
 indicate that the use of the leading-order term only may be insufficient for
 a good description of the $\eta\to 3\pi^0$ decay. Fit no.~2 is shown in
 Fig.~\ref{fig:zeta_eta3pi0_a2_comp}(b) and fit no.~1 in Fig.~\ref{fig:m2pi0_eta3pi0_a2_comp}(b)
 by the cyan long-dashed lines, confirming that it is not sufficient to use only
 the leading-order term. Fit no.~3 in Table~\ref{tab:rslt} was made to the same sextant of
 the raw Dalitz plot with the technique taking both the acceptance and the experimental
 resolution into account (see the text above). This fit results in a slightly better
 $\chi^2$/ndf value and a slightly larger quadratic slope, which was expected because of
 some smearing of the acceptance-corrected distributions by the experimental resolution.
 In the end, the difference between the $\alpha$ results for the acceptance-corrected and
 the raw distributions can be considered as the magnitude of its systematic uncertainty
 due to the limited experimental resolution.

 Fit no.~4 in Table~\ref{tab:rslt}, which also involves the next density-function term 
 $2\beta z^{3/2} \sin(3\phi)$, does improve the $\chi^2$/ndf value, whereas including
 the $2\gamma z^2$ term in fit no.~5 practically does not. In addition, the parameters
 $\alpha$ and $\gamma$ in fit no.~5 become strongly correlated, which results in large fit
 errors for them. Fit no.~4, shown in Figs.~\ref{fig:zeta_eta3pi0_a2_comp}(b)
 and~\ref{fig:m2pi0_eta3pi0_a2_comp}(b) by the yellow solid line, demonstrates
 a quite decent description of the $z$ and $m(\pi^0\pi^0)$ distributions,
 except in the region where the cusp is expected. As shown in the $m(\pi^0\pi^0)$ distribution,
 the $2\beta z^{3/2} \sin(3\phi)$ term curves the spectrum up at the lowest masses,
 which is opposite to the effect expected from the cusp. In the $z$ distribution, the same term
 causes a kink up at $z\approx 0.75$, which again is opposite to the effect expected
 from the cusp~\cite{Prakhov_2009,Rusetsky_2009}.
 As shown in Figs.~\ref{fig:zeta_eta3pi0_a2_comp}(a)
 and~\ref{fig:m2pi0_eta3pi0_a2_comp}(a), the calculation within the framework
 of the modified NREFT~\cite{Schneider_2011} predicts a behavior that is very similar
 to fit no.~4, but with a smaller general slope.
 This can be explained by a smaller quadratic slope, $\alpha=-0.0246(49)$, and positive
 $\gamma=0.0013(4)$ from Ref.~\cite{Schneider_2011}. However, because of the large
 uncertainty in the calculated $\alpha$, it is still in agreement with the corresponding value
 from fit no.~4. In contrast to the calculation from Ref.~\cite{Schneider_2011},
 the prediction based on the extended chiral Khuri-Treiman formalism~\cite{Albaladejo_2017}
 lies below the experimental data points, which is mostly determined by the larger
 quadratic slope, $\alpha=-0.0337(12)$. At the same time, the predictions for
 the $2\beta z^{3/2} \sin(3\phi)$ term, $\beta=-0.0042(7)$~\cite{Schneider_2011}
 and $\beta=-0.0054(1)$~\cite{Albaladejo_2017}, are both in decent
 agreement with the corresponding value from fit no.~4. The experimental value
 for $\gamma$ cannot be determined reliably in order to be compared with the prediction
 from Ref.~\cite{Schneider_2011}.
\begin{table*}
\caption{
 Results from fitting to the acceptance-corrected sextant (ACS), $30^\circ<\phi<90^\circ$, of the
 $\eta\to 3\pi^0$ Dalitz plot with the density function of Eq.~(\protect\ref{eg:azphicu})
 are considered as the main results, and from the other fits as their
 cross checks. The results for the leading-order parametrization
 were also obtained for the acceptance-corrected $z$ (ACZ) distribution and the
 same sextant of the raw (RawS) Dalitz plot.
 The result errors from fitting to the acceptance-corrected half (ACH), $-90^\circ<\phi<90^\circ$,
 of the Dalitz plot are multiplied by the factor of $\sqrt{3}$, correcting for three entries
 per event.
 The results from fitting to the independent data of Run I and Run II
 are added to illustrate systematic effects due to different experimental conditions.
 For convenience, calculations involving the higher-order terms are listed as well.
} \label{tab:rslt}
\begin{ruledtabular}
\begin{tabular}{|c|c|c|c|c|c|c|} 
 Fit no.~ & Data used & $\chi^2$/ndf & $\alpha$ & $\beta$ & $\gamma$ & $\delta$ \\
\hline
 1 & ACS & 1.247 & $-0.0302(8)$ & --- & --- & --- \\
 2 & ACZ & 1.239 & $-0.0304(9)$ & --- & --- & --- \\
 3 & RawS & 1.213 & $-0.0321(9)$ & --- & --- & --- \\
 4 & ACS & 1.119 & $-0.0280(9)$ & $-0.0058(8)$ & --- & --- \\
 5 & ACS & 1.117 & $-0.0231(33)$ & $-0.0053(8)$ & $-0.0057(37)$ & --- \\
 6 & ACS & 1.106 & $-0.0265(10)$ & $-0.0074(10)$ & --- & $-0.0176(68)$ \\
 7 & ACS & 1.108 & $-0.0248(34)$ & $-0.0071(12)$ & $-0.0021(40)$ & $-0.0160(74)$ \\
 8 & ACH & 1.330 & $-0.0302(8)$ & --- & --- & --- \\
 9 & ACH & 1.182 & $-0.0265(10)$ & $-0.0073(10)$ & --- & $-0.0169(67)$ \\
10 & ACH & 1.182 & $-0.0247(33)$ & $-0.0070(12)$ & $-0.0023(40)$ & $-0.0152(73)$ \\
11 & ACS, Run I & 1.212 & $-0.0300(11)$ & --- & --- & --- \\
12 & ACS, Run II & 1.210 & $-0.0304(12)$ & --- & --- & --- \\
13 & ACS, Run I & 1.130 & $-0.0256(15)$ & $-0.0083(14)$ & --- & $-0.0247(95)$ \\
14 & ACS, Run II & 1.154 & $-0.0274(15)$ & $-0.0065(14)$ & --- & $-0.0103(97)$ \\
15 & ACS, Run I & 1.133 & $-0.0246(47)$ & $-0.0081(17)$ & $-0.0013(56)$ & $-0.0237(104)$ \\
16 & ACS, Run II & 1.156 & $-0.0251(48)$ & $-0.0061(17)$ & $-0.0030(58)$ & $-0.0080(106)$ \\
\hline
\hline
 Calculation no.~ & Ref. & --- & $\alpha$ & $\beta$ & $\gamma$ & --- \\
\hline
 1 & \protect\cite{Schneider_2011} & --- & $-0.0246(49)$ & $-0.0042(7)$ & $\gamma=0.0013(4)$ & --- \\
 2 & \protect\cite{Guo_2017} & --- & $-0.025(4)$ & $0.000(2)$ & --- & --- \\
 3 & \protect\cite{Albaladejo_2017} & --- & $-0.0337(12)$ & $-0.0054(1)$ & --- & --- \\
\end{tabular}
\end{ruledtabular}
\end{table*}

 As seen from fit no.~6 in Table~\ref{tab:rslt}, further improvement in the description
 of the $\eta\to 3\pi^0$ data was reached by adding the $2\delta\sum_{i=1}^3\rho(s_i)$
 term, which allows a cusp parametrization to be included in the density function.
 Such a fit results in a slightly smaller quadratic slope, compared to fit no.~4,
 but also in a stronger $2\beta z^{3/2} \sin(3\phi)$ term.
 In Figs.~\ref{fig:zeta_eta3pi0_a2_comp}(b) and~\ref{fig:m2pi0_eta3pi0_a2_comp}(b),
 fit no.~6, which is shown by the green dashed line, demonstrates good agreement with
 both $z$ and $m(\pi^0\pi^0)$ distributions. Based on the results of fit no.~6, 
 the contributions from the $2\beta z^{3/2} \sin(3\phi)$ and the cusp terms partially
 cancel each other in the $z$ and especially in the $m(\pi^0\pi^0)$ distribution.
 Though, according to the result of fit no.~6 for the cusp term,
 the magnitude of the cusp effect at $m(\pi^0\pi^0)=2m_{\pi^0}$ is almost 1\%, its
 visibility here is strongly diminished by the $2\beta z^{3/2} \sin(3\phi)$ term.
 The understanding of such a feature became possible due to fitting the $\eta\to 3\pi^0$
 Dalitz plot based on high experimental statistics.

 The isospin-limit NREFT fit to the present $\eta\to 3\pi^0$ data together with
 KLOE's $\eta\to \pi^+\pi^-\pi^0$ Dalitz plot~\cite{Anastasi_2016} is 
 shown in Figs.~\ref{fig:zeta_eta3pi0_a2_comp}(b)
 and~\ref{fig:m2pi0_eta3pi0_a2_comp}(b) by the blue dash-dotted line.
 As shown in the $m(\pi^0\pi^0)$ distribution, the major deviation of this fit
 from the data is in the cusp region, which is much more prominent in the fit curve.
 The description of the $z$ distribution deviates from the data as well.
 The cusp magnitude obtained at $m(\pi^0\pi^0)=2m_{\pi^0}$ is close to 1\%, which
 is similar to the corresponding result of fit no.~6 in Table~\ref{tab:rslt}.
 The discrepancy seems to come from inability of the isospin-limit fit
 to describe properly the $2\beta z^{3/2} \sin(3\phi)$ term.
 Though the isospin-limit NREFT fit results
 in a good description of the charged decay, with $\chi^2$/ndf=1.072,
 it gives $\chi^2$/ndf=1.290 for the neutral decay.
 The numerical results for $L_i$ were obtained as $L_0 =1(0)$, $L_1 =-4.004(31)$,
 $L_2 =-41.55(31)$, and $L_3 =5.28(14)$, with $K_i$ recalculated from $L_i$ as
 $K_0 =-2.322(7)$ and $K_1 =25.71(73)$.

 The isospin-breaking NREFT fit solely to the present $\eta\to 3\pi^0$ data, which
 is shown in Figs.~\ref{fig:zeta_eta3pi0_a2_comp}(b)
 and~\ref{fig:m2pi0_eta3pi0_a2_comp}(b) by the red dotted line,
 resulted in a much better description of the neutral decay, $\chi^2$/ndf=1.112,
 with the numerical results for $K_i$ and $L_i$ as
 $K_0 =-1.4171(32)$, $K_1 =25.32(29)$, $L_0 =1(0)$, $L_1 =-1.36(44)$,
 $L_2 =-109.5(5.0)$, and $L_3 =-121.0(8.2)$.
 Also, as shown in Figs.~\ref{fig:zeta_eta3pi0_a2_comp}(b)
 and~\ref{fig:m2pi0_eta3pi0_a2_comp}(b), the isospin-breaking NREFT fit
 practically repeats the behavior of fit no.~6 in Table~\ref{tab:rslt},
 which was made with the density function of Eq.~(\protect\ref{eg:azphicu}).

 A comparison of the results from the two NREFT fits indicates a strong
 isospin breaking between the charged and the neutral $\eta\to 3\pi$ decays,
 unless the NREFT framework in Ref.~\cite{Bissegger} could be improved
 for a better simultaneous description of both decay modes. 
 As illustrated in Figs.~\ref{fig:zeta_eta3pi0_a2_comp}(a)
 and~\ref{fig:m2pi0_eta3pi0_a2_comp}(a), a recent dispersive analysis
 by the Bern group~\cite{Colangelo_2017,Heinrich_priv}, in which
 the $\eta\to \pi^+\pi^-\pi^0$ data~\cite{Anastasi_2016}
 were used to determine subtraction constants, did provide predictions
 that described the $\eta\to 3\pi^0$ data well. 
 
 The results of this work provide a strong indication
 that the parametrization of the $\eta\to 3\pi^0$ decay with only the leading-order
 term is insufficient, and the RPP value $\alpha=-0.0318\pm 0.0015$~\cite{PDG}
 reflects a combined effect from higher-order terms and the cusp structure.  
 As the results listed in Table~\ref{tab:rslt} show, the values obtained
 for the quadratic slope parameter become smaller when the higher-order terms
 and the cusp are added, and those values for $\alpha$ are also closer to recent
 calculations reported in Refs.~\cite{Schneider_2011,Guo_2015,Guo_2017}
 (see also Fig.~\ref{fig:alpha_eta3pi0_comp}).

 The exact systematic uncertainties in the results for $\alpha$ and for the other
 parameters are difficult to estimate reliably because the results themselves depend on
 the number of density-function terms included in the fit.
 The systematic effect due to the limited experimental resolution was discussed
 above for a fit with the leading-order term only (no.~3 in Table~\ref{tab:rslt}).  
 The sensitivity of the results to the sextant orientation with respect to
 the plot binning and to additional boundary bins was checked with fits
 to other sextants and to half of the Dalitz plot. All those tests demonstrated
 practically identical results, after multiplying the half-plot errors by
 the factor of $\sqrt{3}$ to correct for three entries per event
 (fits nos.~8---10 in Table~\ref{tab:rslt}).
 The magnitudes of systematic effects for all parameters could also be understood
 by comparing fits to the independent data of Run I and Run II,
 which were taken with different MAMI beam energy and current, target length
 (resulting in different angular resolution), DAQ trigger, energy
 resolution of the calorimeters, etc.
 Those fits are listed as nos.~11---16 in Table~\ref{tab:rslt}.
 As shown, the largest differences between the results from Run I and Run II
 were observed for parameters $\gamma$ and $\delta$; however, all results
 obtained from the different data sets are in agreement within the fit errors.
 The magnitude for parameter $\gamma$ cannot be determined reliably from
 the experimental data because of the large correlation with parameter $\alpha$.
 Therefore, the value obtained for $\alpha$ with the $2\gamma z^2$ term omitted
 actually reflects the combined effect from those two terms. 

 According to the present analysis, the density function of Eq.~(\protect\ref{eg:azphicu})
 with only three parameters is sufficient for a good description of the experimental
 $\eta\to 3\pi^0$ Dalitz plot. The values obtained for these three parameters
 are $\alpha=-0.0265(10_{\rm stat})(9_{\rm syst})$,
 $\beta=-0.0074(10_{\rm stat})(9_{\rm syst})$, and $\delta=-0.018(7_{\rm stat})(7_{\rm syst})$,
 where the main numbers come from fit no.~6 in Table~\ref{tab:rslt}, and the systematic
 uncertainties are taken as half of the differences between the results of fits nos.~13 and 14. 
 The new result for the quadratic slope parameter $\alpha$ strongly indicates that its absolute
 value is smaller by $\approx 20\%$, compared to the previous measurements using
 the leading-order term only. The magnitude of the $2\beta z^{3/2} \sin(3\phi)$ term
 is found to be different from zero by $\sim 5.5$ standard deviations.
 The cusp magnitude obtained at $m(\pi^0\pi^0)=2m_{\pi^0}$ from
 the $2\delta\sum_{i=1}^3\rho(s_i)$ term is close to 1\%, but with an uncertainty
 greater than 50\%. This result is consistent with the prediction for the
 $\eta\to 3\pi^0$ cusp magnitude made within the NREFT model~\cite{Rusetsky_2009}.
  
 The data presented in this work are expected to serve as a valuable input
 for new refined analyses by theoretical groups, which are interested
 in a better understanding of $\eta\to 3\pi$ decays and extracting the quark-mass ratios
 from such data. 

\section{Summary and conclusions}
 The largest, at the moment, statistics of $7\times 10^6$ $\eta\to 3\pi^0$ decays,
 based on $6.2 \times 10^7$ $\eta$ mesons produced in
 the $\gamma p\to \eta p$ reaction,
 has been accumulated by the A2 Collaboration at the Mainz Microtron, MAMI.
 The results of this work provide a strong indication
 that the parametrization of the $\eta\to 3\pi^0$ decay with only the leading-order
 term is insufficient, and the RPP value for $\alpha$
 reflects the combined effect from higher-order terms and the cusp structure,
 whereas the actual quadratic slope is smaller by $\approx 20\%$.
 According to the analysis of the $\eta\to 3\pi^0$ Dalitz plot,
 the cusp magnitude at $m(\pi^0\pi^0)=2m_{\pi^0}$ is about 1\%, but its
 visibility is strongly diminished by the second-order term of the density function,
 the magnitude of which is found to be different from zero by $\sim 5.5$ standard deviations.
 The fits to the present $\eta\to 3\pi^0$ and
 KLOE's $\eta\to \pi^+\pi^-\pi^0$ data within the NREFT framework indicate a strong
 isospin breaking between the charged and the neutral decay modes.
 At the same time, the predictions based on the most recent dispersive analysis
 by the Bern group, in which the $\eta\to \pi^+\pi^-\pi^0$ data
 were used to determine subtraction constants, were found to be in good agreement
 with the present $\eta\to 3\pi^0$ data.
 The data points from the experimental Dalitz plot
 and the ratios of the $z$ and $m(\pi^0\pi^0)$ distributions to phase space
 are provided as supplemental material to the paperl~\cite{Supplemental}. 

\begin{acknowledgments}
 The authors acknowledge the excellent support of the accelerator group and
 operators of MAMI.
 We thank H. Leutwyler, G. Colangelo, and B. Kubis for fruitful
 discussions and constant interest in our work.
 This work was supported by the Deutsche Forschungsgemeinschaft (SFB443,
 SFB/TR16, and SFB1044), DFG-RFBR (Grant No. 09-02-91330), the European Community-Research
 Infrastructure Activity under the FP6 ``Structuring the European Research Area''
 program (Hadron Physics, Contract No. RII3-CT-2004-506078), Schweizerischer
 Nationalfonds (Contracts No. 200020-156983, No. 132799, No. 121781, No. 117601, No. 113511),
 the U.K. Science and Technology Facilities Council (STFC 57071/1, 50727/1),
the U.S. Department of Energy (Offices of Science and Nuclear Physics,
 Awards No. DE-FG02-99-ER41110, No. DE-FG02-88ER40415, No. DE-FG02-01-ER41194)
 and National Science Foundation (Grants No. PHY-1039130, No. IIA-1358175),
 INFN (Italy), and NSERC of Canada (Grant No. FRN-SAPPJ-2015-00023).
 A.~Fix acknowledges additional support from the Tomsk Polytechnic University
 competitiveness enhancement program.
 We thank the undergraduate students from Department of Physics of Mount Allison University
 and from Institute for Nuclear Studies of The George Washington University for their assistance.
\end{acknowledgments}


\begin{thebibliography}{99}

\bibitem{Gasser} J. Gasser and H. Leutwyler,
                 Nucl. Phys. B {\bf 250}, 539 (1985).

\bibitem{Bijnens_2002} J.~Bijnens and J.~Gasser, 
                 Phys.\ Scripta\ T {\bf 99}, 034 (2002).

\bibitem{Bijnens_2007} J.~Bijnens and K.~Ghorbani, 
                 JHEP\ {\bf 11}, 030 (2007).

\bibitem{Kambor} J. Kambor {\it et al.},
                 Nucl. Phys. B {\bf 465}, 215 (1996).

\bibitem{anisovich} A. Anisovich and H. Leutwyler,
                 Phys. Lett. B {\bf 375}, 335 (1996).

\bibitem{Colangelo_2001} G.~Colangelo, J.~Gasser, and H.~Leutwyler,
  Nucl.\ Phys.\ B\ {\bf 603}, 125 (2001).

\bibitem{Kaminski_2008} R.~Kami\'nski, J.~R.~Pel\'aez, and F.~J.~Yndur\'ain,
 Phys.\ Rev.\ D\ {\bf 77}, 054015 (2008),
 
\bibitem{Garcia_2011} R.~Garc\'ia-Mart\'in, R.~Kami\'nski, J.~R.~Pel\'aez, J.~Ruiz de Elvira,
 and F.~J.~Yndur\'ain, Phys.\ Rev.\ D\ {\bf 83}, 074004 (2011),

\bibitem{eta_slope_bnl} W.~B.~Tippens {\it et al.}, 
    Phys. Rev. Lett. {\bf 87}, 192001 (2001).

\bibitem{Unverzagt_2009} M.~Unverzagt {\it et al.},
              Eur. Phys. J. A {\bf 39}, 169 (2009)

\bibitem{Prakhov_2009} S.~Prakhov {\it et al.},
 Phys.\ Rev.\ C\ {\bf 79}, 035204 (2009).

\bibitem{Ambrosino_2008} F.~Ambrosino {\it et al.},
  JHEP {\bf 05}, 006 (2008), 

\bibitem{Adlarson_2014} P.~Adlarson {\it et al.},
  Phys.\ Rev.\ C\ {\bf 90}, 045207 (2014)

\bibitem{Anastasi_2016} A.~Anastasi {\it et al.},
  JHEP {\bf 05}, 019 (2016).

\bibitem{Rusetsky_2009} C.~O.~Gullstr\"om, A.~Kup\'s\'c, and A.~Rusetsky,
 Phys.\ Rev.\ C\ {\bf 79}, 028201 (2009).

\bibitem{Schneider_2011} S.~P.~Schneider, B.~Kubis, and C.~Ditsche,
 JHEP {\bf 02}, 028 (2011).

\bibitem{Kampf_2011} K.~Kampf, M.~Knecht, J.~Novotn\'y, and M.~Zdr\'ahal,
 Phys.\ Rev.\ D\ {\bf 84}, 114015 (2011).

\bibitem{Guo_2015} P.~Guo, I.~V.~Danilkin, D.~Schott, C.~Fern\'andez-Ram\'irez,
V.~Mathieu, and A.~P.~Szczepaniak,
 Phys.\ Rev.\ D\ {\bf 92}, 054016 (2015).

\bibitem{Guo_2017} P.~Guo, I.~V.~Danilkin, D.~Schott, C.~Fern\'andez-Ram\'irez,
V.~Mathieu, and A.~P.~Szczepaniak,
 Phys.\ Lett.\ B\ {\bf 771}, 497 (2017).

\bibitem{Kolesar_2017} M.~Koles\'ar and J.~Novotn\'y,
 Eur.\ Phys.\ J.\ C\ {\bf 77}, 41 (2017).

\bibitem{Colangelo_2017} G.~Colangelo, S.~Lanz, H.~Leutwyler, and E.~Passemar,
 Phys.\ Rev.\ Lett.\ {\bf 118}, 022001 (2017).

\bibitem{Albaladejo_2017} M.~Albaladejo and B.~Moussallam,
 Eur.\ Phys.\ J.\ C\ {\bf 77}, 508 (2017).

\bibitem{PDG} C.~Patrignani {\it et al.}, (Particle Data Group),
 Chin.\ Phys.\ C\ {\bf 40}, 100001 (2016).

\bibitem{GAMS_1984} D.~Alde {\it et al.},
  Z.\ Phys.\ C\ {\bf 25}, 225 (1984).

\bibitem{CBarrel_1998} A.~Abele {\it et al.},
  Phys.\ Lett.\ B\ {\bf 417}, 193 (1998).

\bibitem{SND_2001} M.~N.~Achasov {\it et al.},
 JETP\ Lett.\ {\bf 73}, 451 (2001).

\bibitem{WASA_2007} M.~Bashkanov {\it et al.},
 Phys.\ Rev.\ C\ {\bf 76}, 048201 (2007).

\bibitem{WASA_2009} C.~Adolph {\it et al.},
 Phys.\ Lett.\ B\ {\bf 677}, 24 (2009).

\bibitem{KLOE_2011} F.~Ambrosino {\it et al.},
 Phys.\ Lett.\ B\ {\bf 694}, 16 (2011).

\bibitem{BESIII_2015} M.~Ablikim {\it et al.},
 Phys.\ Rev.\ D\ {\bf 92}, 012014 (2015).

\bibitem{Borasoy} B.~Borasoy and R.~Nissler,
                  Eur. Phys. J. A {\bf 26}, 383 (2005). 

\bibitem{NA48_2009} J.~R.~Batley {\it et al.},
  Eur.\ Phys.\ J.\ C\ {\bf 64}, 589 (2009).

\bibitem{Heinrich_priv} Heinrich Leutwyler, private communication.

\bibitem{Bissegger} M.~Bissegger, A.~Fuhrer, J.~Gasser, B.~Kubis
                 and A.~Rusetsky,
                 Phys.\ Lett.\ B\ {\bf 659}, 576 (2008).

\bibitem{A2_eta_etapr_2017} V.~L.~Kashevarov {\it et al.},
 Phys.\ Rev.\ Lett.\ {\bf 118}, 212001 (2017).

\bibitem{CB} A.~Starostin {\it et al.},
 Phys.\ Rev.\ C\ {\bf 64}, 055205 (2001).

\bibitem{TAPS} R.~Novotny,  
  IEEE Trans.\ Nucl.\ Sci.\ {\bf 38}, 379 (1991).

\bibitem{TAPS2} A.~R.~Gabler {\it et al.},
 Nucl.\ Instrum.\ Methods\ Phys.\ Res.\ A\ {\bf 346}, 168 (1994).

\bibitem{MAMI} H.~Herminghaus {\it et al.},
         IEEE Trans.\ Nucl.\ Sci.\ {\bf 30}, 3274 (1983).

\bibitem{MAMIC} K.-H.~Kaiser {\it et al.},
  Nucl.\ Instrum.\ Methods\ Phys.\ Res.\ A\ {\bf 593}, 159 (2008).

\bibitem{TAGGER} I.~Anthony {\it et al.},
  Nucl.\ Instrum.\ Methods\ Phys.\ Res.\ A\ {\bf 301}, 230 (1991).

\bibitem{TAGGER1} S.~J.~Hall {\it et al.},
  Nucl.\ Instrum.\ Methods\ Phys.\ Res.\ A\ {\bf 368}, 698 (1996).

\bibitem{TAGGER2} J.~C.~McGeorge {\it et al.},
  Eur.\ Phys.\ J.\ A\ {\bf 37}, 129 (2008).

\bibitem{etamamic} E.~F.~McNicoll {\it et al.},
 Phys.\ Rev.\ C\ {\bf 82}, 035208 (2010).

\bibitem{PID} D.~Watts, {\it Proceedings of the 11th International
              Conference on Calorimetry in Particle Physics},
             Perugia, Italy, 2004 (World Scientific, Singapore, 2005), p. 560.

\bibitem{k0sn_lpi0_spi0_2009} S.~Prakhov {\it et al.},
        Phys.\ Rev.\ C\ {\bf 80}, 025204 (2009).

\bibitem{K0Sigpl2013} P.~Aguar-Bartolom\'e {\it et al.},
  Phys.\ Rev.\ C\ {\bf 88}, 044601 (2013).

%\bibitem{Supplemental} See Supplemental Material at [URL] for the data points from the experimental
% Dalitz plot and the ratios of the $z$ and $m(\pi^0\pi^0)$ distributions to phase space.
\bibitem{Supplemental} See Supplemental Material for the data points from the experimental
 Dalitz plot and the ratios of the $z$ and $m(\pi^0\pi^0)$ distributions to phase space.

\end{thebibliography}
\end{document}